\documentstyle[psfig,12pt,axodraw]{article}
\begin{document}
\baselineskip 18pt
\def\today{\ifcase\month\or
 January\or February\or March\or April\or May\or June\or
 July\or August\or September\or October\or November\or December\fi
 \space\number\day, \number\year}
\def\thebibliography#1{\section*{References\markboth
 {References}{References}}\list
 {[\arabic{enumi}]}{\settowidth\labelwidth{[#1]}
 \leftmargin\labelwidth
 \advance\leftmargin\labelsep
 \usecounter{enumi}}
 \def\newblock{\hskip .11em plus .33em minus .07em}
 \sloppy
 \sfcode`\.=1000\relax}
\let\endthebibliography=\endlist
\begin{titlepage}
%
%\hspace*{10.0cm}ICRR-Report-405-97-28

\hspace*{10.0cm}OCHA-PP-118
\  \
\vskip 2.0 true cm
\renewcommand{\thefootnote}
{\fnsymbol{footnote}}
\begin{center} 
\large {\bf Decay Rate Asymmetry of the Top Quark \\
       through Light Squarks
\footnote{Talk given at the Workshop on Fermion Mass and CP violation,
Hiroshima, Japan, 5-6 March 1998.}
}
\vskip 2.0 true cm
Mayumi Aoki \footnote{Research Fellow of the Japan Society
for the Promotion of Science.}\\
\vskip 0.5 true cm  
{\it Graduate School of Humanities and Sciences, Ochanomizu University \\
Otsuka 2-1-1, Bunkyo-ku, Tokyo 112-8610, Japan}
\end{center}

\vskip 4.0 true cm

\centerline{\bf Abstract}
\medskip

     A decay rate asymmetry of the top quark is discussed
within the framework of 
the supersymmetric standard model.
Although new sources of $CP$ violation in this model are severely 
constrained from the electric dipole moments
of the neutron and the electron, 
there is a possibility that a top squark has a mass of around
100 GeV and 
$CP$ violation in its interaction
with a top quark is unsuppressed. 
Then partial widths could be different
between the decays $t\rightarrow bW^+$
and $\bar t\rightarrow \bar bW^-$.
The magnitude of this $CP$ asymmetry can be of 
order $10^{-3}$, 
which may be detectable in 
the near future.

\medskip

\end{titlepage}

\def\gsim{{\mathop >\limits_\sim}}
\def\lsim{{\mathop <\limits_\sim}}
\def\r2{\sqrt 2}
\def\a#1{\alpha_#1}
\def\sw2{\sin^2\theta_W}
\def\tw{\tan\theta_W}
\def\v#1{v_#1}
\def\tb{\tan\beta}
\def\c2b{\cos 2\beta}
\def\w{\omega}
\def\x{\chi}
\def\g{\tilde g}
\def\sq{\tilde q}
\def\su{\tilde u}
\def\sd{\tilde d}
\def\st{\tilde t}
\def\sb{\tilde b}
\def\sl{\tilde l}
\def\m#1{{\tilde m}_#1}
\def\mg{{\tilde m}_g}
\def\mH{m_H}
\def\mgr{m_{3/2}}
\def\mw#1{\tilde m_{\omega #1}}
\def\mx#1{\tilde m_{\chi #1}}
\def\M{\tilde M}

\section{Introduction}
In quantum field theory,
$CPT$ invariance holds under only fundamental assumptions,
while $CP$ violation can be implemented by various ways.
Phenomenologically, $CP$ violation is observed 
in $K^0-\bar K^0$ mixing, and possibly in the universe as baryon
asymmetry.
Although the former is described well by the Kobayashi-Maskawa 
mechanism of the standard model (SM),
it is difficult to produce observed baryon asymmetry 
through this mechanism.
It seems necessary to extend the SM to have new sources
of $CP$ violation.

One plausible extension of the SM is 
the supersymmetric
standard model (SSM) based on $N$=1 supergravity coupled to grand unified
theories (GUTs).
This model contains new sources of $CP$ violation,
which could generate baryon asymmetry sufficiently $\cite{aoki}$.
The electric dipole moments (EDMs) of the neutron and the electron
are also predicted to have large magnitudes $\cite{edm}$.
However, the new sources of $CP$ violation
do not affect much the $K^0-\bar K^0$ and $B^0-\bar B^0$ systems.
The SSM can only contribute to $CP$ asymmetries in $B$-meson decays
indirectly through  $B^0-\bar B^0$ mixing $\cite{branco}$.
Other observables have to be invoked for examining $CP$ violation
by the SSM in collider experiments. 

In this paper, I report a recent study by Aoki and Oshimo $\cite{tdecay}$ on 
$CP$ violation induced by the SSM in 
the production and decay of the top quark.
If one top squark is light enough, a top quark can decay into
a top squark and a neutralino, which subsequently become 
a bottom quark and a $W$ boson
through final state interactions.
The interaction of a top squark with a top quark could violate
$CP$ invariance maximally.
Then, $CP$ violation may be observed as a difference of
partial decay rates between the decays of a top
quark and an anti-top quark, which is measured by
the decay rate asymmetry 
\begin{equation}
A_{CP}=\frac{\Gamma(t\rightarrow bW^+)-\Gamma(\bar t\rightarrow \bar bW^-)}
       {\Gamma(t\rightarrow bW^+)+\Gamma(\bar t\rightarrow \bar bW^-)}.
\end{equation}
Taking into account the constraints from the EDMs of the neutron and 
the electron, we discuss $A_{CP}$ and explore parameter region
where a sizable asymmetry is expected.

\section{The model}
The SSM has several complex parameters.
In the supersymmetric part of the interaction Lagrangian,
complex are a mass parameter $m_H$ in the linear coupling of Higgs
superfields and coefficients of the trilinear Yukawa couplings.
The soft-breaking part contains gaugino mass parameters
$\m3, \m2$, and $\m1$ for the SU(3),
SU(2), and U(1) gauge groups, respectively,
dimensionless coupling constants $A_f$'s and $B$
for the trilinear and linear terms
of scalar fields, respectively.
We assume unification of the gaugino masses at the GUT scale,
giving the relation $\m3/\a3=\m2/\a2=3\m1/5\a1$ at the electroweak scale.
Since $A_f$'s 
are considered to  
have the same value of order unity
at the GUT scale, their differences at the electroweak scale
are neglected and thus we put $A_f=A$.
Under these assumptions, by redefining 
particle fields, we can take $m_H$ and $A$
as physical complex parameters 
without loss of generality.
We express these parameter as $m_H=|m_H|\exp(i\theta)$ and 
$A=|A|\exp(i\alpha)$.
The phase $\theta$ is contained in the chargino and neutralino mass matrices
and the squark and slepton mass-squared matrices.
The phase $\alpha$ is contained only in the latter.

     The mass-squared matrix $M^2_q$ for the squarks corresponding to
a quark $q$ with mass $m_q$, electric charge $Q_q$,
and third component of the weak isospin $T_{3q}$ is given by
\[
    \lefteqn{M^2_q =} \hspace{9cm}
\]
\[
 \left(\matrix{m_q^2 + \c2b (T_{3q} - Q_q\sw2 )M_Z^2 + \M_{qL}^2 &
                                            m_q (R_q\mH + A^*\mgr) \cr
                   m_q (R_q\mH^* + A\mgr) &
                               m_q^2 +  Q_q\c2b\sw2 M_Z^2 + \M_{qR}^2}
           \right),
\]
\begin{eqnarray}
   R_q &= & \frac{1}{\tb} \quad (\ T_{3q} = \frac{1}{2}\ ),
    \quad  \tb \quad (\ T_{3q} = -\frac{1}{2}\ ), \nonumber
\label{sqmass} \\
   \tb &= & \frac{\v2}{\v1}, 
\end{eqnarray}
where $\M_{qL}^2$ and $\M_{qR}^2$ denote
the mass-squared parameters for the left-handed squark
and the right-handed squark, respectively, and
$\mgr$ is the gravitino mass.
Generation mixings are not necessary to induce $CP$ violation, so that
we neglect them.
The mass eigenstates $\sq_1$ and $\sq_2$ are obtained by diagonalizing
the mass-squared matrix as
\begin{equation}
      S_q^\dagger\tilde M_q^2 S_q = {\rm diag}(\M_{q1}^2, \M_{q2}^2)
\quad
                                    (\M_{q1}^2<\M_{q2}^2),
\end{equation}
where $S_q$ is a unitary matrix.
The slepton mass-squared
matrices are obtained by appropriately changing $M_q^2$ in
Eq. (\ref{sqmass}).

At the GUT scale, we consider the squark and slepton masses to have
a common value of the gravitino mass.
Then, at the electroweak scale, the values of the mass-squared parameters 
for the squarks of
the first two generations and all the sleptons are approximately the same,
\begin{equation}
\M_{qL}^2 \simeq \M_{qR}^2 \simeq \M_{lL}^2 \simeq \M_{lR}^2 \equiv \M^2.
\end{equation}
On the other hand, those for the squarks of the third generation
receive large quantum corrections through Yukawa interactions proportional
to the top quark mass $m_t$, expressed as
\begin{eqnarray}
 \M_{tL}^2 &=& \M^2-cm_t^2,  \quad \M_{tR}^2 = \M^2-2cm_t^2,
    \nonumber \\
 \M_{bL}^2 &=& \M^2-cm_t^2,  \quad \M_{bR}^2 = \M^2,
\end{eqnarray}
with $c = 0.1-1$.
Under this scheme, if $\tilde M$ is around $m_t$, the quantum 
corrections and the large values of the off-diagonal elements of 
the top squark mass-squared matrix make 
one top squark rather light.

In our scheme, the SSM parameters which determine the interactions at 
the electroweak scale are $\tan\beta$, $A$, $m_H$, $\tilde m_2$, 
$\tilde M$, $m_{3/2}$, and c.
Although these parameters are not all independent each other,
they can have various sets of values depending on assumptions
for underlying GUTs and parameter values.
Therefore, for simplicity, we take those parameters independent.

\section{Constraints from EDMs}
The complex mass matrices for the $R$-odd particles lead to
$CP$-violating interactions, which give rise to the EDMs of 
the neutron and the electron at the one-loop level.
The exchanged particles in the loop diagrams are charginos, neutralionos,
and gluinos with squarks or sleptons.
The present experimental upper bounds on the neutron and the electron EDMs
are approximately $10^{-25}e$.cm and $10^{-26}e$.cm, respectively.
For an unsuppressed value of the $CP$-violating phase $\theta$, 
both the neutron and the electron EDMs generally receive
dominant contributions from the chargino-loop diagrams,
which are approximately
proportional to $\sin\theta$.
The experimental bounds on the EDMs impose the constraints that
the squarks and the sleptons should be heavier than 1 TeV
$\cite{edm}$.
The chargino contributions become small as the magnitude of
$\theta$ decreases, irrespective of the value of another
$CP$-violating phase $\alpha$.
For a sufficiently small value of $\theta$, the constraints
on the squark and the slepton masses are relaxed.

Assuming $\theta \ll 1$ and $\alpha \sim 1$,
the neutron and
the electron EDMs receive dominant contributions from the gluino-
and the neutralino-loop diagrams, respectively.
In Table 1, the absolute values of the EDMs of the neutron 
$|d_n|$ and the electron $|d_e|$ are shown
for several values of the SU(2) gaugino mass $\tilde m_2$,
taking $\M =200$ GeV, $\theta =0$, and $\alpha =\pi/4$.
The other parameters are taken as tan$\beta=2, |m_H|=100$ GeV, and
$|A|m_{3/2}=\tilde M$.
If $\tilde m_2$ is around or greater than 500 GeV,
the neutron EDM is consistent with its experimental bound.
In all the range of $\tilde m_2$, the predicted value
of the electron EDM lies within the experimental bound. 
This numerical analysis shows that
even if the phase $\alpha$ is of order unity, 
squarks and sleptons are allowed to have masses of order 100 GeV, 
without causing inconsistency for the EDMs of the neutron and the
electron.
\begin{table}
\caption{The absolute values of the EDMs of the neutron $|d_n|$
and the electron $|d_e|$ for several different values
of $\m2$.
The $CP$-violating phases are taken as  $\alpha=\pi/4$ and $\theta=0$.
We have chosen  $\tb=2$, $|\mH|=100$ GeV,
and $|A|\mgr=\M=200$ GeV. }
\vspace{1cm}
\begin{center}
\begin{tabular}{ccc}
     $\m2$ (GeV) & $|d_n|$ & $|d_e|$  \\
\hline
  300 & $3.4\times 10^{-25}$ & $1.3\times 10^{-26}$   \\
  400 & $1.8\times 10^{-25}$ & $1.1\times 10^{-26}$   \\
  500 & $1.1\times 10^{-25}$ & $9.5\times 10^{-27}$   \\
  600 & $7.5\times 10^{-26}$ & $8.1\times 10^{-27}$  \\
  700 & $5.2\times 10^{-26}$ & $6.9\times 10^{-27}$  \\
  800 & $3.8\times 10^{-26}$ & $5.8\times 10^{-27}$  \\
  900 & $2.9\times 10^{-26}$ & $5.0\times 10^{-27}$  \\
\end{tabular}
\end{center}
\end{table}

\section{Numerical results of decay rate asymmetry}
The constraints from the EDMs do not 
exclude a possibility that the lighter top squark
is lighter than the top quark.
There is a parameter region where a top quark can decay
into a top squark and a neutralino.
The produced top squark and neutralino can yield a bottom quark
and a $W$ boson by   
exchanging charginos or bottom squarks as shown in Fig. 1.
If $CP$ invariance is violated in these reactions,
the interference of the decay amplitudes at the tree level and the one-loop
level makes the partial decay rates different between the decays
$t \rightarrow bW^+$ and $\bar t \rightarrow \bar bW^-$.
The decay rate asymmetry $A_{CP}$ is obtained as
\begin{eqnarray}
A_{CP} &=& \frac{\alpha_2}{2}\left[\biggl\{m_t^2+m_b^2-2M_W^2+
          \frac{(m_t^2-m_b^2)^2}{M_W^2}\biggl\}
    \sqrt{\lambda(m_t^2,M_W^2,m_b^2)}\right]^{-1} T,
\end{eqnarray}
where $T$ consists of $CP$-odd terms 
coming from the contributions of the diagrams (a) and (b) in Fig. 1.
We refer its analytical formula to Ref. $\cite{tdecay}$.
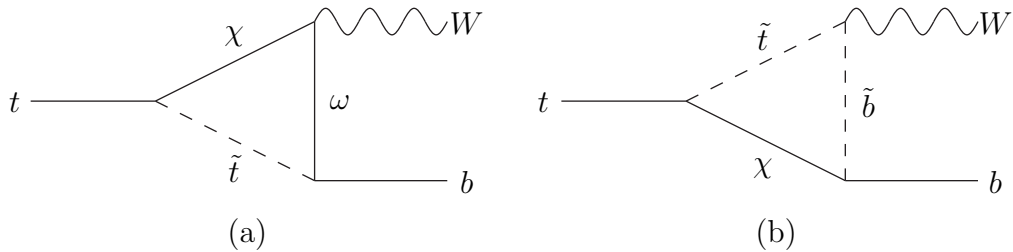
\begin{figure}
\caption{The one-loop diagrams for the top quark decay
producing a bottom quark and a $W$ boson,
where $\st$, $\sb$, $\w$, and $\x$ denote respectively
the top squark, bottom squark, chargino, and neutralino.
  }
\label{oneloop}
\vspace{2cm}
%
%\unitlength=1mm
%\SetScale{2}
\begin{center}\begin{picture}(400,100)(0,0)
\Line(23,100)(70,100)
\Line (70,100)(130,130)
\DashLine (70,100)(130,70){5}
\Line(130,130)(130,70)
\Photon(130,130)(180,130){5}{3}
\Line(130,70)(180,70)
\Text(17,100)[]{$t$}
\Text(100,125)[]{$\chi$}
\Text(100,75)[]{$\tilde t$}
\Text(140,100)[]{$\omega$}
\Text(188,130)[]{$W$}
\Text(188,70)[]{$b$}
\Text(105,50)[]{(a)}
%%%%
\Line(223,100)(270,100)
\DashLine (270,100)(330,130){5}
\Line (270,100)(330,70)
\DashLine(330,130)(330,70){5}
\Photon(330,130)(380,130){5}{3}
\Line(330,70)(380,70)
\Text(217,100)[]{$t$}
\Text(300,125)[]{$\tilde t$}
\Text(300,75)[]{$\chi$}
\Text(340,100)[]{$\tilde b$}
\Text(388,130)[]{$W$}
\Text(388,70)[]{$b$}
\Text(305,50)[]{(b)}
\end{picture}
\end{center}
\end{figure}

We give numerical results of the absolute value of the $CP$ asymmetry 
$|A_{CP}|$ in Table 2
for $\theta=0$, $\alpha=\pi/4$ and 
several different values of $\m2$,
taking three sets of values for $\tilde M$ and $c$ :
(i)$\M=180$ GeV, $c=0.3$, (ii)$\M=200$ GeV, $c=0.4$, 
(iii)$\M=220$ GeV, $c=0.5$.
The other parameters are taken as $\tan \beta=2, |m_H|=100$ GeV, and
$|A|m_{3/2}=\tilde M$.
In these parameter regions
the experimental constraints on
the squark and chargino masses are satisfied.
The top quark mass is taken to be 180 GeV.
Resultant values of the lighter top squark mass are given by
(i) 97 GeV, (ii) 92 GeV, (iii) 91 GeV.
For larger values of $\tilde m_2$ 
in Table 2 for which no value of $A_{CP}$ is presented, 
the decay $t \rightarrow 
\tilde t \chi$ is not allowed kinematically.

The numerical analysis shows that, for $\theta \ll 1$,  $\alpha \sim 1$,
$\tilde M \sim$ 200 GeV, and $m_H \sim$ 100 GeV,
the absolute value of the $CP$ asymmetry $A_{CP}$ 
is of order $10^{-3}$ and maximally $2 \times 10^{-3}$.
This amount of asymmetry 
can be detected if the decays $t \rightarrow bW^+$ and $ \bar t
\rightarrow \bar b W^-$ are found by an amount of order $10^6$.
At hadron colliders, it is not so easy to find these decays by 
purely hadronic final states, so that an isolated lepton coming from
the $W$ boson decay may be used as a signal of the top quark decay.
In the parameter space where the asymmetry $A_{CP}$ has 
a sizable value, the branching ratio for $t
\rightarrow bW^+$ is around 0.8.
Since the branching ratio for the leptonic decay of
a $W$ boson is around 0.2, 
the production of $t\bar t$ pairs of order $10^7$ will enable
the detection of the decay rate asymmetry.
At the CERN Large Hadron Collider
(LHC), the top quark pairs are expected to be produced at a rate 
of order $10^7$.
The asymmetry may be detectable in near-future experiments
at LHC.

\begin{table}
\caption{The decay rate asymmetry $A_{CP}$ is shown 
   for several different values of $\m2$. 
We take the $CP$-violating
phases for $\alpha=\pi/4$ and $\theta=0$.
The other parameters are taken as $\tb=2$, $|\mH|=100$ GeV,
  and $|A|\mgr=\M$.
We have chosen three sets of values for $\M$ and $c$ :
(i)$\M=180$ GeV, $c=0.3$, (ii)$\M=200$ GeV, $c=0.4$, 
(iii)$\M=220$ GeV, $c=0.5$.
  }
\vspace{0.8cm}
\begin{center}
\begin{tabular}{c|c|c|c|}
 $\m2$ (GeV) & 
\multicolumn{3}{c|}{$|A_{CP}|$} \\
\cline{2-4} & (i)& (ii) & (iii) \\
\hline
  300  & $1.7\times 10^{-3}$ & $1.9\times 10^{-3}$ & $1.9\times 10^{-3}$ \\
  400  & $1.7\times 10^{-3}$ & $2.0\times 10^{-3}$ & $2.1\times 10^{-3}$\\
  500  & $1.4\times 10^{-3}$ & $1.8\times 10^{-3}$ & $2.0\times 10^{-3}$\\    
  600  & $7.7\times 10^{-4}$ & $1.6\times 10^{-3}$ & $1.7\times 10^{-3}$  \\
  700  &                     & $1.2\times 10^{-3}$ & $1.4\times 10^{-3}$  \\
  800  &                   &   $8.0\times 10^{-4}$ & $1.0\times 10^{-3}$  \\
  900  &                   &                   &  $7.0\times 10^{-4}$ \\
\end{tabular}
\end{center}
\end{table}

The interactions which induce the rate asymmetry
between the standard decays $t \rightarrow bW^+$ and $\bar t 
\rightarrow \bar b W^-$
also yield a rate asymmetry between the non-standard decays
$t \rightarrow \tilde t \chi$ and $\bar t \rightarrow
\tilde t^* \chi$. 
This difference, however, compensates the difference for the standard 
decays, satisfying the relation
\begin{equation}
\Gamma(t\rightarrow bW^+)-\Gamma(\bar t\rightarrow \bar bW^-)=
-\left\{\Gamma(t\rightarrow \st\x)-\Gamma(\bar t\rightarrow
\st^*\x)\right\}.
\end{equation}
As a result, the total width of the top quark is the same as that
of the anti-top quark, as required by $CPT$ invariance.
On the other hand, this relation makes the detection of the decay rate
asymmetry more involved.
In most of the parameter region where the asymmetry 
is saizable, the top squark decays through $\st_1 \rightarrow
b\w^+ \rightarrow bW^*\x$,
where $W^*$ denotes the virtual state of the $W$ boson.
Therefore, 
the final states of the non-standard decay and the standard
decay have the same particles except invisible neutralinos.
In order to measure the asymmetry, therefore, detailed analysis of
final states is needed to 
distinguish between the decays $t \rightarrow bW^+$
and $t \rightarrow \tilde t \x$.
In addition, top squark pairs are directly produced at a rate larger
than the top quark pairs.
Since the decays $t\rightarrow bW^+$ and $\tilde t\rightarrow b\omega^+$
yield the same visible particles, we also
need to distinguish between these decays.
Such distinctions can be made by examining energy
spectra of the particles in the final states.

We have seen a possibility that $CP$ violation is observed at the
electroweak energy scale if the decay $t\rightarrow \st \chi$
is allowed kinematically.
Such a non-standard decay, however, may be detected at Tevatron, 
if it has a large branching ratio.
The branching ratio of $t\rightarrow \tilde t_1 \chi_1$ is 
approximately 0.2.
Since the final states of the top quark decay and top squark decay
have the same visible particles, as mentioned above, 
energy spectra of these particles
have to be examined in detail to find the non-standard decay mode.
At present, it is still not ruled out that the non-standard decay may 
have escaped detection, even if its branching ratio is comparable
to that of the standard decay.    

Through our discussions, we have assumed the masses of the squarks
and the sleptons to be degenerate at the GUT scale.
By some unknown reasons, however, it may happen that
the squarks of the third generation are of order 100 GeV
and not related to the mass of the other 
squarks and sleptons.
In this case, if the squarks and sleptons of the first generation are
heavier than 1 TeV, the EDMs of the neutron and the electron
do not impose constraints on
the $CP$-violating phases.
We have also computed the $CP$ asymmetry taking both $\theta$ and $\alpha$
of order unity.
The asymmetry is at most approximately $3\times 10^{-3}$ and
not significantly large compared to the asymmetry obtained under 
the assumption of unification for the squark and slepton masses.

\section{Summary}
We have discussed a partial decay rate asymmetry between
the two decays $t \rightarrow bW^+$ and $\bar t \rightarrow \bar b W^-$
within the framework of the SSM.
As long as one $CP$-violating phase $\theta$ is sufficiently small,
even if another phase $\alpha$ is of order unity, 
squarks, sleptons, one chargino, and some of neutralinos are allowed
to have masses of order 100 GeV without inconsistency
with the constraints from the EDMs of the neutron and the electron.
In this case there is a parameter space where
the non-standard decay $t \rightarrow \tilde t \chi$
is allowed kinematically. 
In a wide region of this parameter space, the asymmetry is of order
$10^{-3}$ and maximally $2\times 10^{-3}$.
Since $10^7$ pairs of top quarks are expected to be produced at LHC,
such an amount of asymmetry may be detectable
in the near future. 

The lighter top squark may be heavier than the top quark.
In this case, the decay rate asymmetry $A_{CP}$ in Eq. (1)
cannot be used to measure $CP$ violation.
However, if the top squark is sufficiently heavy, another rate
asymmetry between the decays $\tilde t \rightarrow t\x$ and
$\tilde t^* \rightarrow \bar t \x$ is induced by the same sources of $CP$
violation in the SSM $\cite{stdecay}$.
The parameter space where this asymmetry has a non-vanishing value
is wide, compared to that for $A_{CP}$
which is restricted because of 
the kinematical constraint $m_t > \M_1 +\tilde m_{\chi_1}$.
The investigation of the decay rate asymmetry for the top squark
and anti-top squark is
another window for $CP$ violation in the SSM.

\section*{Acknowledgments}
The author enjoyed the fruitful collaboration with Noriyuki Oshimo,
whom she thanks for reading the manuscript.
This work is supported in part by the Grant-in-Aid for
Scientific
Research from the Ministry of Education, Science and Culture, Japan.

\end{document}